\documentclass[twocolumn,aps,prd,eqsecnum,showpacs,preprintnumbers]{revtex4}
\usepackage{graphicx}
\begin{document}
\preprint{KUNS:1958}
\title
{Dynamical bar instability in a relativistic rotational core collapse}
%
\author{Motoyuki Saijo}
\email{Motoyuki.Saijo@obspm.fr}
\altaffiliation
{\\Present address:
Laboratoire de l'Univers et de ses Th\'eories,\\
Observatoire de Paris, F-92195 Meudon Cedex,  France}
%
\affiliation
{Department of Physics, Kyoto University,
Kyoto 606-8502, Japan}
%
\received{28 February 2005}
\published{31 May 2005}
%
\begin{abstract}
We investigate the rotational core collapse of a rapidly rotating relativistic 
star by means of a 3+1 hydrodynamical simulations in conformally flat 
spacetime of general relativity.  We concentrate our investigation to the 
bounce of the rotational core collapse,  since potentially most of the 
gravitational waves from it are radiated around the core bounce.  The dynamics 
of the star is started from a differentially rotating equilibrium star of 
$T/W \sim 0.16$ ($T$ is the rotational kinetic energy and $W$ is the 
gravitational binding energy of the equilibrium star), depleting the pressure 
to initiate the collapse and to exceed the threshold of dynamical bar 
instability.  Our finding is that the collapsing star potentially forms a bar 
when the star has a toroidal structure due to the redistribution of the 
angular momentum at the core bounce. On the other hand, the collapsing star 
weakly forms a bar when the star has a spheroidal structure.  We also find 
that the bar structure of the star is destroyed when the torus is destroyed in 
the rotational core collapse.  Since the collapse of a toroidal star 
potentially forms a bar, it can be a promising source of gravitational waves 
which will be detected in advanced LIGO.
\end{abstract}
%
\pacs{04.25.Dm,04.30.Db,04.40.Dg,97.60.-s}
\maketitle
%
\section{Introduction}

Rotational core collapse occurs in the first phase of the collapse-driven
supernova explosion.  The collapse takes place in a dynamical time scale and 
the bounce occurs when the central part of the star collapses into a nuclear 
density and then stiffens its equation of state.  The final outcome of the 
collapse forms a compact object, such as a neutron star or a black hole, 
which is a candidate for sources of gravitational waves \citep{CT02}.  There 
are two different approaches to investigate the rotational core collapse.  One 
is to take into account the realistic picture of the rotational collapse such 
as realistic equation of state of the neutron star and/or neutrino effect in 
Newtonian gravity \citep[e.g.][]{FH00,KYS03,MRBJS}.  The other is to 
illustrate simplified physics in relativistic gravitation 
\citep[e.g.][]{DFM01,DFM02}.  Since one of our aims in this paper is to 
investigate the possibility of gravitational wave sources in rotational core 
collapse, we should at least take into account relativistic gravitation, 
which significantly affects the quantitative behavior of gravitational waves 
from rotational core collapse \citep{DFDFIMS}.

Dynamical bar instability in a rotating equilibrium star takes place when the 
ratio $\beta (\equiv T/W)$ between rotational kinetic energy $T$ and the 
gravitational binding energy $W$ exceeds the critical value $\beta_{\rm dyn}$. 
Determining the onset of the dynamical bar-mode instability, as well as the 
subsequent evolution of an unstable star, requires a fully nonlinear 
hydrodynamic simulation.  Simulations performed in Newtonian gravity
\citep[e.g.][]{TDM,DGTB,WT,HCS,SHC,HC,TIPD,NCT,LL,Liu02} have shown that 
$\beta_{\rm dyn}$ depends only very weakly on the stiffness of the equation of 
state.  $\beta_{\rm dyn}$ becomes small for stars with high degree of 
differential rotation \citep{TH90,PDD,SKE}.  Simulations in relativistic 
gravitation \citep{SBS00,SSBS01} have shown that $\beta_{\rm dyn}$ decreases 
with the compaction of the star, indicating that relativistic gravitation 
enhances the bar-mode instability.  The dynamical bar instability potentially 
occurs during the collapse since the nondimensional value $\beta$ scales as 
$R^{-1}$ in the dimensional analysis where $R$ is the radius of the star.  Let 
us briefly discuss a picture of core bounce.  When the star collapses, $\beta$ 
increases and the star starts to deform its shape to form a bar when the star 
exceeds the critical value of dynamical instability.  During the bounce phase, 
$\beta$ falls below the threshold of dynamical instability, and the star 
cannot deform furthermore.  In such a situation, what happens to the 
nonaxisymmetric deformation of the star?

There are several papers that investigate rotational core collapse as a 
potential source of gravitational waves in axisymmetric spacetime in Newtonian 
gravity \citep{ZM,KYS03}, in conformally flat spacetime 
\citep{DFM01,DFM02,Siebel03}, and in full general relativity 
\citep{SAF,SS04,Shibata04} to estimate the amount of gravitational radiation 
\citep{FHH02}.  Recently, 3D calculations in full general relativity have been 
established to investigate the nonaxisymmetric deformation of the star in 
rotational core collapse.  \citet{DSY} investigated the collapse of a 
differentially rotating $n=1$ polytropic star in 3D by depleting the pressure 
and found that the collapsing star forms a torus which fragments into 
nonaxisymmetric clumps.  \citet{SS05} investigated in rotational core collapse 
and found that a burst type of gravitational waves was emitted.  In addition, 
they argued that a very limited window for the rotating star satisfies to 
exceed the threshold of dynamical instability in the core collapse.  
\citet{Zink05} presented a fragmentation of an $n=3$ toroidal polytropic star 
to a binary system inducing $m=2$ density perturbation and claimed that they 
found a new scenario to form a binary black hole.

Our purpose in this paper is the following twofold.  One is to investigate a 
necessary condition and mechanism to enhance the dynamical bar instability in 
a collapsing star.  \citet{Brown} performed long time integration of the 
rotational core collapse to investigate whether a dynamical bar can 
significantly form during the evolution.  He found that the dynamical bar 
instability sets in at $\beta_{\rm dyn} \approx 0.23$, which is far below the 
standard value $\beta_{\rm dyn} \approx 0.27$.  He found that the role of 
dynamical bar in rotational core collapse may be quite different from that in 
the equilibrium star.  He also found that the bar instability grows slowly in 
core bounce.  His finding can be interpreted that bar instability is initiated 
by the interaction between the core part and the surrounding part of the star.
Therefore the ``dynamical'' bar instability in the collapsing star is quite 
different from that in the equilibrium state.

The other is the importance of probing whether the rotational core collapse 
becomes a promising source of gravitational waves.  Direct detection of 
gravitational waves by ground based and space based interferometers is of 
great importance in general relativity, in astrophysics, and in cosmology.  
Once a bar has formed in the neutron star, we may expect quasi-periodic 
gravitational waves in the {\rm kHz} band, which may be detectable in advanced 
LIGO \citep{CT02}.  \citet{RMR} investigated the rotational core collapse 
using parametric equation of state in Newtonian gravity, including the thermal 
pressure and the nuclear matter when the density exceeds the threshold of 
nuclear density.  They found that although the rotating star has a 
nonaxisymmetric deformation, the amplitude of gravitational waves does not 
significantly grow during the deformation.  In fact, they compared the 
gravitational waveform computed in their simulation with that in the previous 
2D calculation \citep{ZM} and found that it is quite similar to each other.  
Therefore, even if the nonaxisymmetric instability takes place during the 
collapse, the bar does not significantly grow at the core bounce.  What else 
do we need to enhance the bar formation in the rotational core collapse? 

Our interest is to focus on the core bounce of the rotational core collapse 
in order to investigate the angular momentum redistribution and to investigate 
the possibility of gravitational wave sources.  Accordingly we construct a 
mimic model to investigate the above two issues.  We deplete the pressure of 
the equilibrium star to initiate collapse and bounce.  We also concentrate on 
the structure of the star, spheroidal and toroidal, to investigate the 
dynamical bar instability in the collapsing star.  Stellar collapses and 
mergers may also lead to differentially rotating stars \citep[e.g.][]{Ott04}.  
For the coalescence of binary irrotational neutron stars 
\citep{SU00,SU02,STU03}, the presence of differential rotation may temporarily 
stabilize the ``hypermassive'' remnant, which constructs a toroidal structure 
and may therefore have important dynamical effects.  Although they use stiff 
equation of state ($n=1$) it is possible to construct a toroidal star in 
nature.

This paper is organized as follows.  In Sec. \ref{sec:bequation} we summarize 
our basic equation of relativistic hydrodynamics in conformally flat 
spacetime.  In Sec. \ref{sec:NR} we show our numerical results of dynamical 
bar instability in rotational core collapse, and summarize our findings in 
Sec. \ref{sec:Discussion}.  Throughout this paper we use geometrized units 
($G = c = 1$) and adopt Cartesian coordinates $(x,y,z)$ with the coordinate 
time $t$.  Greek and Latin indices run over $(t, x, y, z)$ and $(x, y, z)$, 
respectively.

\section{Relativistic Hydrodynamics in Conformally Flat Spacetime}
\label{sec:bequation}

In this section, we describe the basic equations in conformally flat spacetime 
\citep[e.g.][]{IN,WM89,Saijo04}.  We solve the fully relativistic equations of 
hydrodynamics, but neglect nondiagonal spatial metric components.

\subsection{The gravitational field equations}

We define the spatial projection tensor $\gamma^{\mu\nu} \equiv g^{\mu\nu} + 
n^{\mu} n^{\nu}$, where $g^{\mu\nu}$ is the spacetime metric, $n^{\mu} = 
(1/\alpha, -\beta^i/\alpha)$ the unit normal to a spatial hypersurface, and 
where $\alpha$ and $\beta^i$ are the lapse and shift.  Within a first 
post-Newtonian approximation, the spatial metric $g_{ij} = \gamma_{ij}$ may 
always be chosen to be conformally flat
\begin{eqnarray}
\gamma_{ij} = \psi^{4} \delta_{ij},
\end{eqnarray}
where $\psi$ is the conformal factor \citep[see][]{Chandra65, BDS}.  The 
spacetime line element then reduces to
\begin{eqnarray}
&&
ds^{2} =
( - \alpha^{2} + \beta_{k} \beta^{k} ) dt^{2} + 2 \beta_{i} dx^{i} dt 
+ \psi^{4} \delta_{ij} dx^{i} dx^{j}.
\nonumber
\\
\end{eqnarray}
We adopt maximal slicing, for which the trace of the extrinsic curvature 
$K_{ij}$ vanishes,
\begin{equation}
K \equiv \gamma^{ij} K_{ij} = 0.
\end{equation}  
The gravitational field equations in conformally flat spacetime for the five 
unknown $\alpha$, $\beta^i$, and $\psi$ can then be derived conveniently 
from the 3+1  formalism.  The equation for the lapse $\alpha$, shift 
$\beta^{i}$, and conformal factor $\psi$ with a maximal slicing implies 
$\partial_t K = 0$, shall be written as
\begin{widetext}
\begin{eqnarray}
&&
\triangle (\alpha \psi) = 2 \pi \alpha \psi^{5}
(\rho_{\rm H} + 2 S) + \frac{7}{8} \alpha \psi^{5} K_{ij} K^{ij},
\\
&&
\delta_{il} \triangle \beta^{l} +
\frac{1}{3} \partial_{i} \partial_{l} \beta^{l} =16 \pi \alpha J_{i}
+
\left(\partial_{j}\ln \left( \frac{\alpha}{\psi^{6}} \right)\right)
\left(
   \partial_{i} \beta^{j} + \delta_{il} \delta^{jk} \partial_{k} \beta^{l} -
  \frac{2}{3} \delta_{i}^{j} \partial_{l} \beta^{l}
\right),
\\
&&
\triangle \psi =
- 2 \pi \psi^{5} \rho_{\rm H} - \frac{1}{8} \psi^{5} K_{ij} K^{ij},
\label{eqn:HamiltonianC}
\end{eqnarray}
\end{widetext}
where $S  = \gamma_{jk} T^{jk}$, $\Delta \equiv \delta^{ij} \partial_i 
\partial_j$ is the flat space Laplacian and $J_{i} \equiv -n^{\mu} 
\gamma^{\nu}_{~i} T_{\mu \nu}$ is the momentum density.  In the definition of 
$J_{i}$, $T_{\mu\nu}$ is the stress energy tensor, $\rho_{\rm H} \equiv 
n^{\mu} n^{\nu} T_{\mu \nu}$ is the mass-energy density measured by a normal 
observer.

\subsection{The matter equations}
For a perfect fluid, the energy momentum tensor takes the form 
\begin{equation}
T^{\mu \nu} = 
\rho \left( 1 + \varepsilon + \frac{P}{\rho} \right) u^{\mu} u^{\nu} +
Pg^{\mu\nu},
\end{equation}
where $\rho$ is the rest-mass density, $\varepsilon$ the specific internal 
energy, $P$ the pressure, and $u^{\mu}$ the four-velocity.  We adopt a 
$\Gamma$-law equation of state in the form
\begin{equation}
P = (\Gamma - 1) \rho \varepsilon,
\label{eqn:gammalaw1}
\end{equation}
where $\Gamma$ is the adiabatic index which we set $\Gamma = 5/3$, $3/2$, 
$7/5$ in this paper.  In the absence of thermal dissipation, 
Eq.~(\ref{eqn:gammalaw1}), together with the first law of thermodynamics, 
implies a polytropic equation of state
\begin{equation}
P = \kappa \rho^{1+1/n},
\end{equation}
where $n=1/(\Gamma-1)$ is the polytropic index and $\kappa$ is a constant.  

From $\nabla_{\mu} T^{\mu\nu}=0$ together with the equation of state (Eq. 
[\ref{eqn:gammalaw1}]), we can derive the energy and Euler equations according 
to
\begin{widetext}
\begin{eqnarray}
&&
\frac{\partial e_{*}}{\partial t}+
\frac{\partial (e_{*} v^{j})}{\partial x^{j}} = 
- \frac{1}{\Gamma}(\rho \epsilon)^{-1+1/\Gamma} 
P_{\rm vis}
\frac{\partial}{\partial x^{i}} 
( \alpha u^{t} \psi^{6} v^{i} )
\label{eqn:Energy}
,\\
&&
\frac{\partial(\rho_{*} \tilde u_{i})}{\partial t}
+ \frac{\partial (\rho_* \tilde u_{i} v^{j})}{\partial x^{j}} 
=
- \alpha \psi^{6} (P + P_{\rm vis})_{,i} 
- \rho_{*} \alpha \tilde u^{t} \alpha_{,i} 
+ \rho_{*} \tilde u_{j} \beta^{j}_{~,i}
+ \frac{2 \rho_{*} \tilde u_{k} \tilde u_{k}}{\psi^{5} \tilde u^{t}} 
\psi_{,i}
,  
\label{eqn:Euler}
\end{eqnarray}
\end{widetext}
where 
\begin{eqnarray}
e_{*} &=& (\rho \varepsilon)^{1/\Gamma} \alpha u^{t} \psi^{6},\\
v^{i} &=& {dx^i \over dt}=\frac{u^{i}}{u^{t}},\\
\rho_{*} &=& \rho \alpha u^{t} \psi^{6},\\
\tilde{u}^{t} &=& ( 1 + \Gamma \varepsilon ) u^{t}
,\\
\tilde{u}_{i} &=& ( 1 + \Gamma \varepsilon ) u_{i}
,
\end{eqnarray}
and $v^{i}$, $u^{\mu}$, $P_{\rm vis}$ is the 3-velocity, 4-velocity, pressure 
viscosity, respectively.  Note that we treat the matter fully 
relativistically; the conformally flat approximation only enters through 
simplifications in the coupling to the gravitational fields.  As a consequence 
to treat shocks we also need to solve the continuity equation
\begin{equation}
\frac{\partial \rho_{*}}{\partial t}
+\frac{\partial (\rho_{*} v^{i})}{\partial x^{i}} = 0,
\label{eqn:BConservation}
\end{equation}
separately.

\begin{table*}
\caption
{Relativistic Rotating Equilibrium stars}
\begin{center}
\begin{tabular}{c c c c c c c c c c c}
\hline
\hline
Model &
$n$\footnotemark[1] &
$\hat{A}$\footnotemark[2] &
$R_{p}/R_{e}$\footnotemark[3] &
$\rho_{0}^{\rm max}$\footnotemark[4] &
$R_{c}$\footnotemark[5] &
$M$\footnotemark[6] &
$J$\footnotemark[7] &
$T/W$\footnotemark[8] &
$R_{c}/M$ &
$P_{\rm dep}$\footnotemark[9]
\\
\hline
I-a & $1.5 $ & $0.5$ & $0.467$ &
$8.26 \times 10^{-4}$ & $6.56$ & 
$1.64 \times 10^{-1}$ & $4.49 \times 10^{-2}$ & 
$1.59 \times 10^{-1}$ & $3.99 \times 10^{1}$ &
$70 \%$
\\
I-b & $1.5$ & $0.8$ & $0.467$ &
$1.36 \times 10^{-3}$ & $7.22$ & 
$1.79 \times 10^{-1}$ & $5.25 \times 10^{-2}$ & 
$1.59 \times 10^{-1}$ & $4.04 \times 10^{1}$ &
$70 \%$
\\
\hline
II-a & $2.0$ & $0.5$ & $0.450$ &
$1.10 \times 10^{-4}$ & $2.53 \times 10^{1}$ & 
$6.29 \times 10^{-1}$ & $6.33 \times 10^{-1}$ & 
$1.63 \times 10^{-1}$ & $4.03 \times 10^{1}$ &
$60 \%$
\\
II-b & $2.0$ & $0.6$ & $0.417$ &
$2.19 \times 10^{-4}$ & $2.64 \times 10^{1}$ & 
$6.62 \times 10^{-1}$ & $6.69 \times 10^{-1}$ & 
$1.62 \times 10^{-1}$ & $3.98 \times 10^{1}$ &
$60 \%$
\\
\hline
III-a & $2.5$ & $0.3$ & $0.417$ &
$1.05 \times 10^{-5}$ & $8.63 \times 10^{1}$ & 
$2.16$ & $6.90$ & 
$1.59 \times 10^{-1}$ & $3.99 \times 10^{1}$ &
$50 \%$
\\
III-b & $2.5$ & $0.42$ & $0.417$ &
$2.24 \times 10^{-5}$ & $8.54 \times 10^{1}$ & 
$2.13$ & $6.58$ & 
$1.60 \times 10^{-1}$ & $4.00 \times 10^{1}$ &
$50 \%$
\\
\hline
\end{tabular}
\label{tbl:initial}
\footnotetext[1]{Polytropic index}
\footnotetext[2]{Parameter of the degree of differential rotation}
\footnotetext[3]{Ratio of the polar proper radius to the 
equatorial proper radius}
\footnotetext[4]{Maximum rest-mass density}
\footnotetext[5]{Equatorial circumferential radius}
\footnotetext[6]{Gravitational mass}
\footnotetext[7]{Angular momentum}
\footnotetext[8]{$T$: Rotational kinetic energy; 
$W$: Gravitational binding energy}
\footnotetext[9]{Ratio of pressure depletion to the equilibrium star}
\end{center}
\end{table*}

\subsection{Numerical techniques for solving gravitational field equations}

We have reduced Einstein equations in conformally flat spacetime to ten 
elliptic equations for ten variables ($B_{i}$, $\chi$, $\psi$, $\alpha \psi$, 
$P_{i}$, $\eta$) using the same techniques in the previous paper 
\citep{Saijo04}
\begin{eqnarray}
\Delta B_{i} &=& 8 \pi \psi^{6} J_{i} \equiv 4 \pi S_{B_{i}},
\label{eqn:CFBx}
\\
\Delta \chi  &=& - 8 \pi \psi^{6} J_{i} x^{i} \equiv 4 \pi S_{\chi},
\\
\Delta \psi  &=& 
- 2 \pi \psi^{5} \rho_{\rm H} - 
\frac{1}{8} \psi^{-7} \hat{A}_{ij} \hat{A}^{ij}
\equiv 4 \pi S_{\psi},
\\
\Delta (\alpha \psi) &=& 
2 \pi \alpha \psi (\rho_{\rm H} + 2 S) +
\frac{7}{8} \alpha \psi^{-7} \hat{A}_{ij} \hat{A}^{ij}
\nonumber \\
&\equiv& 4\pi S_{\alpha\psi}, 
\\
\Delta P_{i} &=& 4 \pi \alpha \hat{J}_{i} \equiv 4 \pi S_{P_{i}},
\\
\Delta \eta  &=& -4 \pi \alpha \hat{J}_{i} x^{i} \equiv 4 \pi S_{\eta}
\label{eqn:CFeta}
.
\end{eqnarray}

We use the asymptotic fall off behavior for metric quantities at large radius 
in order to set an appropriate boundary condition at the grid edge 
\citep{Saijo04}.  The definition of $B_{i}$, $\chi$, $P_{i}$, $\eta$ and the 
basic procedure to solve ten elliptic equations are given in Ref. 
\citep{Saijo04}.  Since we parallelized our code, we change the method to 
solve the elliptic equation to PCG method \citep[e.g.][]{PCG}.

We monitor the rest mass $M_{0}$, gravitational mass $M$, and the angular 
momentum $J$ 
\begin{eqnarray}
M_{0} &=& \int \rho_{*} d^{3}x
,\\
M &=& - \frac{1}{2 \pi} \oint_{\infty} \nabla^{i} \psi dS_{i}
\nonumber \\
&=&
\int 
\biggl[ 
  \left[ 
    ( \rho + \rho \varepsilon + P ) (\alpha u^{t})^{2} - P 
  \right]
  \psi^{5} 
\nonumber \\
&&
+ \frac{1}{16 \pi} \psi^{5} K_{ij} K^{ij}
\biggl] d^{3}x
,\\
J &=&
- \frac{1}{2 \pi} \oint_{\infty} ( x K^{j}_{y} - y K^{j}_{x} ) \psi^{6} dS_{i}
\nonumber \\
&=&
\int (x J_{y} - y J_{x}) \psi^{6} d^{3} x
,
\end{eqnarray}
during the evolution.  We also compute rotational kinetic energy $T$, proper 
mass $M_{p}$, gravitational binding energy $W$ as
\begin{eqnarray}
M_{p} &=&
\int \rho u^{t} ( 1 + \epsilon ) \sqrt{-g} d^{3} x
\nonumber \\
&=&
\int \rho_{*} ( 1 + \varepsilon ) d^{3} x
,\\
T &=&
\frac{1}{2} \int \Omega T^{t}_{\phi} \sqrt{-g} d^{3} x 
\nonumber \\
&=& \frac{1}{2} \int \Omega (x J_{y} - y J_{x}) \psi^{6} d^{3} x
,\\
W &=&
M_{p} + T - M,
\end{eqnarray}
where $\Omega$ the angular velocity of the star.

Since we use a polytropic equation of state at $t=0$, it is convenient to
rescale all quantities with respect to $\kappa$.  Since $\kappa^{n/2}$ has 
dimensions of length, we introduce the following nondimensional variables 
\citep[e.g.][]{CST92}
\begin{equation}
\begin{array}{c c c}
\bar{t} = \kappa^{-n/2} t
, &
\bar{x} = \kappa^{-n/2} x
, &
\bar{y} = \kappa^{-n/2} y
, \\
\bar{z} = \kappa^{-n/2} z
, &
\bar{\Omega} = \kappa^{n/2} \Omega
, &
\bar{M} = \kappa^{-n/2} M
, \\
\bar{R} = \kappa^{-n/2} R
, &
\bar{J} = \kappa^{-n} J
. &
\end{array}
\end{equation}
Henceforth, we adopt nondimensional quantities, but omit the bars for
convenience (equivalently, we set $\kappa = 1$).

\begin{figure*}
\begin{center}
\includegraphics[width=0.7\textwidth]{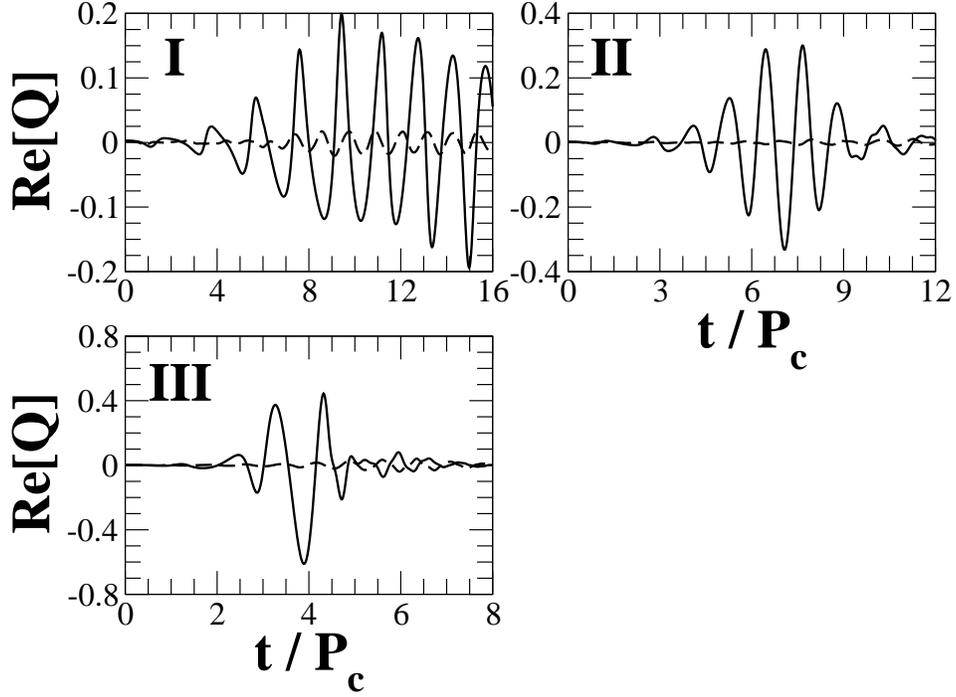}
\end{center}
\caption{
Evolution of quadrupole diagnostics in 6 different collapsing stars.  Roman 
character in the panel corresponds to the model type in Table 
\ref{tbl:initial}.  Solid lines and dash lines represent toroidal and 
spheroidal stars, which correspond to model $a$ and $b$ in Table 
\ref{tbl:initial}, respectively. 
\label{fig:dig}
}
\end{figure*}

\begin{figure*}
\begin{center}
\includegraphics[width=0.7\textwidth]{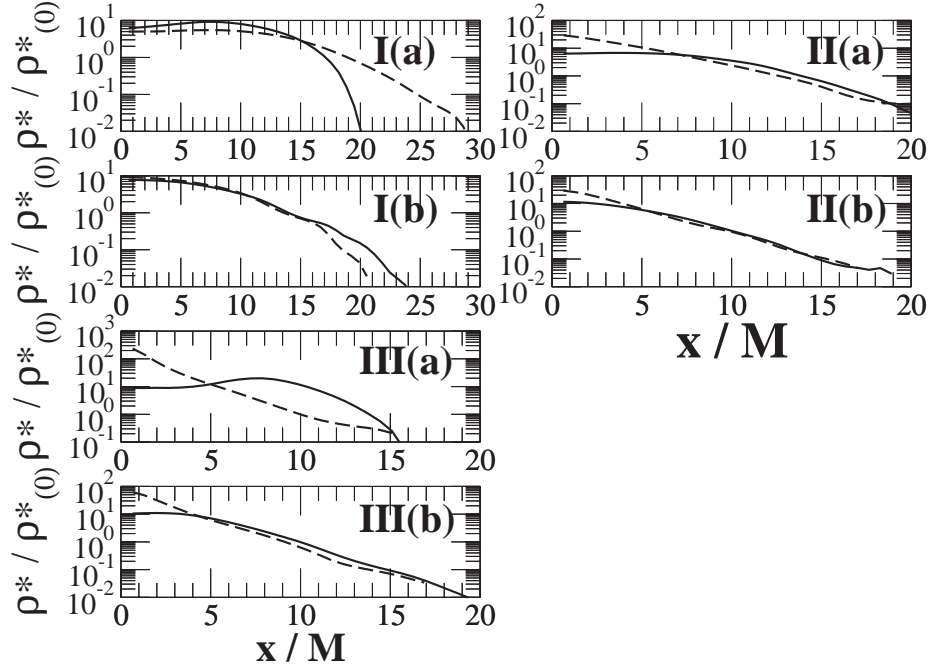}
\end{center}
\caption{
Snapshot of the coordinate density along the $x$-axis for 6 different stars 
(see Table \ref{tbl:initial}).  Solid lines and dash lines represent 
$t/P_{\rm c}=$ 
  I(a) (9.61, 16.0),   I(b) (9.67, 15.3), 
 II(a) (2.39, 12.0),  II(b) (3.25, 12.0), 
III(a) (3.19, 7.77), III(b) (3.20, 7.77), respectively.
\label{fig:density}
}
\end{figure*}

\begin{figure*}
\begin{center}
\includegraphics[width=0.7\textwidth]{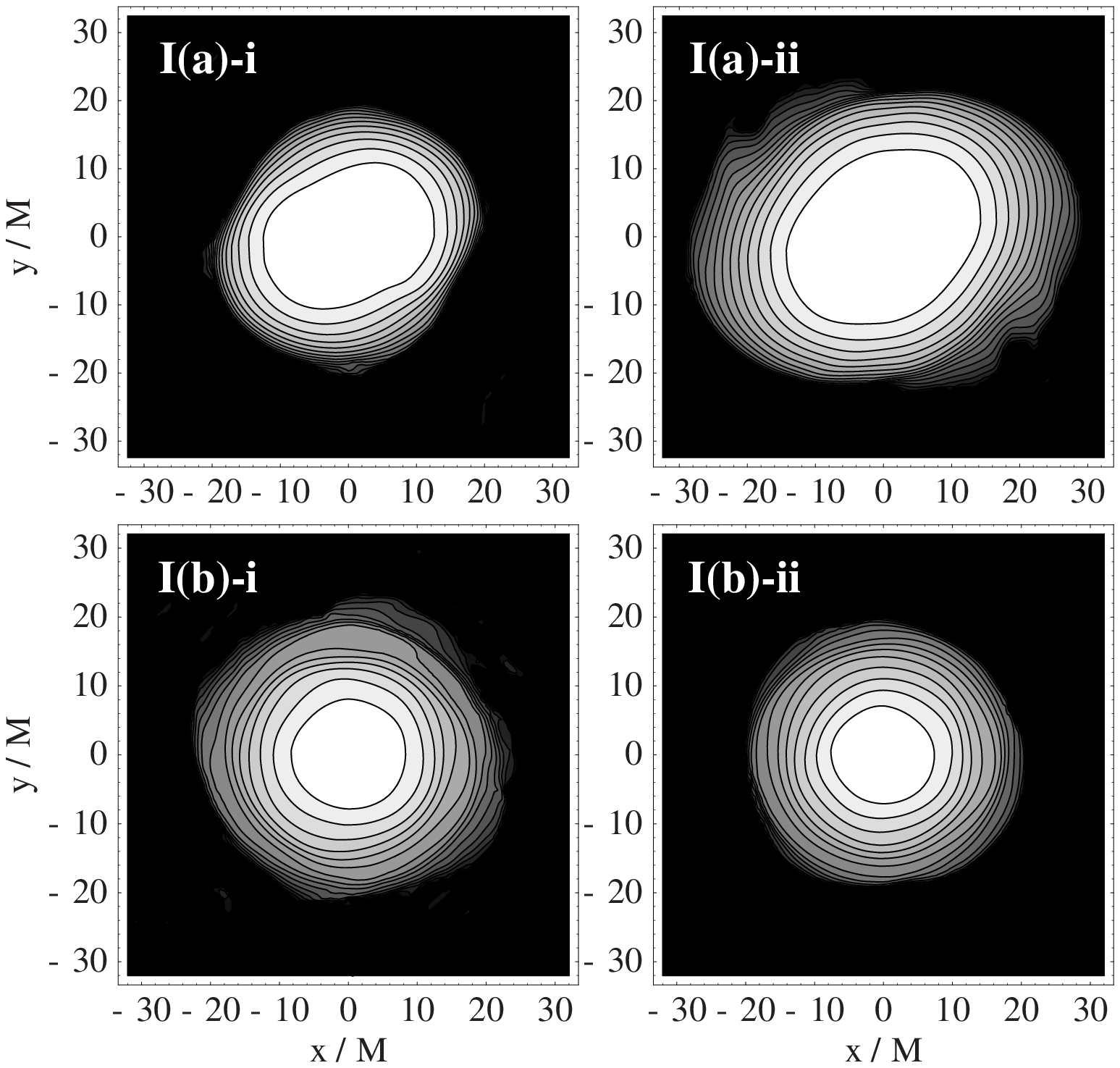}
\end{center}
\caption{
Density contour in the equatorial plane of 2 collapsing stars (Model I[a] and
[b] of Table \ref{tbl:initial}).  Snapshots are plotted at the parameter 
($t/t_{\rm dyn}$, $\rho^{*}_{\rm max}$) $=$ 
I(a)-i ($9.61$, $9.13 \times 10^{-3}$), 
I(a)-ii ($16.0$, $6.04 \times 10^{-3}$), 
I(b)-i ($9.67$, $1.33 \times 10^{-2}$), 
I(b)-ii ($15.3$, $1.57 \times 10^{-2}$), 
respectively.  The contour lines denote coordinate densities
$\rho^{*} = \rho^{*}_{\rm max} \times 10^{- 0.220 (16-i)}  (i=1, \cdots, 15)$.
\label{fig:n15con}
}
\end{figure*}

Our code is based on the conformally flat hydrodynamics scheme of
Ref. \cite{Saijo04}, to which the reader is referred for a more detailed 
description, discussion and test.  We choose the axis of rotation to align 
with the $z$ axis, and assume planar symmetry across the equator.  The 
equations of hydrodynamics are then solved on a uniform grid of size $200 
\times 200 \times 60$.  We terminate our simulations after a sufficient
number of central rotation periods (between 8 and 16) in order for us to 
detect dynamical instabilities.  Because of our flux-conserving difference 
scheme the rest mass $M_{0}$ is conserved up to a round-off error, except if 
matter leaves the computational grid (which was never more than 0.01\% of the 
total rest mass).  In all cases reported in Sec.~\ref{sec:NR} the total 
gravitational mass $M$ and the angular momentum $J$ were conserved up to 
$\sim 0.1 \%$ and less than about $5 \%$ of their initial values.

\section{Rotational Core Collapse}
\label{sec:NR}
We basically follow the scheme of Ref. \citep{KEH89} to construct a 
differentially rotating equilibrium star.  The detailed procedure we used 
for constructing it is given in Ref. \citep{Saijo04}.  We construct the 
equilibrium star under the condition of fixing $\beta$ and $R_{\rm c}/M$,
where $R_{\rm c}$ is the circumferential radius.  This is because we keep 
approximately the same condition to extend the threshold of dynamical bar 
instability by depleting the pressure of the star to initiate collapse.  We 
set one toroidal star (corresponds to model a in Table \ref{tbl:initial}) and 
one spheroidal star (corresponds to model b in Table \ref{tbl:initial}) for 
each polytropic index to focus on the structure of the star.

\begin{figure*}
\begin{center}
\includegraphics[width=0.7\textwidth]{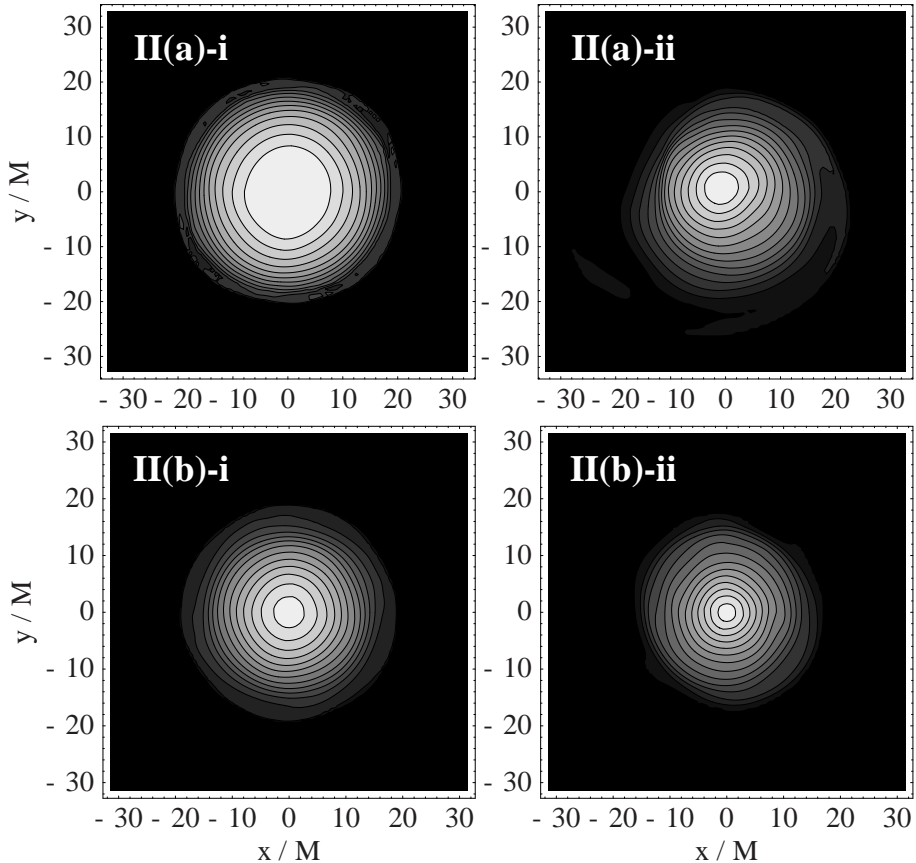}
\end{center}
\caption{
Density contour in the equatorial plane of 2 collapsing stars (Model II[a] and
[b] of Table \ref{tbl:initial}).  Snapshots are plotted at the parameter 
($t/t_{\rm dyn}$, $\rho^{*}_{\rm max}$) $=$ 
II(a)-i ($3.19$, $1.13 \times 10^{-3}$), 
II(a)-ii ($12.0$, $4.62 \times 10^{-3}$), 
II(b)-i ($3.25$, $3.39 \times 10^{-3}$), 
II(b)-ii ($12.0$, $8.71 \times 10^{-3}$), respectively.
The contour lines denote coordinate densities
$\rho^{*} = \rho^{*}_{\rm max} \times 10^{- 0.200 (16-i)}  (i=1, \cdots, 15)$.
\label{fig:n20con}
}
\end{figure*}

We choose the rotation profile of the star as \citep{KEH89}
\begin{equation}
u^{t} u_{\varphi} = A^2 (\Omega_{c} - \Omega)
,
\end{equation}
which represents in the Newtonian limit ($u^{t} \rightarrow 1$, 
$u_{\varphi} \rightarrow \varpi^2 \Omega$), so-called $j$-constant law, as
\begin{equation}
\Omega = \frac{A^{2} \Omega_{\rm c}}{\varpi^{2} + A^{2}}
,
\end{equation}
where $A$ is a parameter representing the degree of differential rotation, 
$\varpi$ is the cylindrical radius of the star.  Since $A$ has a dimension of 
length, we normalize it with the proper equatorial radius $\bar{R}_{e}$, 
($A = \bar{R}_{e} \hat{A}$).  We summarize our 6 different initial data sets 
of relativistic rotating equilibrium stars in Table \ref{tbl:initial}.

To monitor the development of $m=2$ modes we compute a ``quadrupole 
diagnostic'' \citep{SBS03}
\begin{equation}
Q = \left< e^{i m \varphi} \right>_{m=2} =
  \frac{1}{M_{0}} \int \rho_{*} 
  \frac{(x^{2}-y^{2}) + i (2 x y)}{x^{2}+y^{2}} d^3 x,
\label{eqn:quadrupole}
\end{equation}
where a bracket denotes the density weighted average.  In the following we 
only plot the real parts of $Q$.

To enhance any $m=2$ instability, we disturb the initial equilibrium density 
$\rho_{\rm eq}$ by a nonaxisymmetric perturbation:
\begin{equation}
\rho = \rho_{\rm eq}
\left( 1 +
  \delta \frac{x^{2}-y^{2}}{R_{\rm eq}^{2}}
\right),
\label{eqn:DPerturb}
\end{equation}
with $\delta = 0.01$ in all our simulations.

As for computing the gravitational waveform, we use the same method that we 
used in the previous paper \citep{SSBS01}.  For observers along the rotational 
axis ($z$-axis), we have
\begin{eqnarray}
\frac{r h_{+}}{M} &=&
\frac{1}{2 M} \frac{d^{2}}{d t^{2}} (I_{xx} - I_{yy}), \label{h+}
\\
\frac{r h_{\times}}{M} &=&
\frac{1}{M} \frac{d^{2}}{d t^{2}} I_{xy} \label{h-}
,
\end{eqnarray}
where $h_{+}$ and $h_{\times}$ are the two polarization modes of gravitational
waves, $r$ is the distance to the source, $I_{ij}$ the approximate quadrupole
moment of the mass distribution defined as
\begin{equation}
I_{ij} = \int \rho_{*} x^{i} x^{j} d^{3}x.
\end{equation}

\begin{figure*}
\begin{center}
\includegraphics[width=0.7\textwidth]{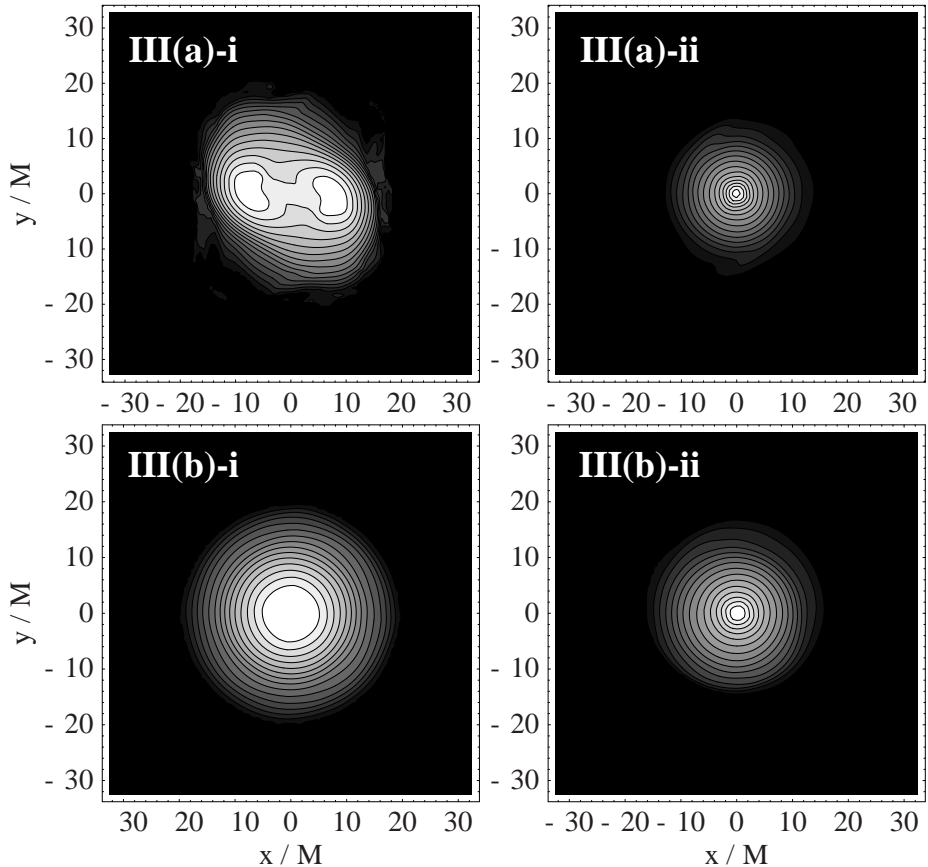}
\end{center}
\caption{
Density contour in the equatorial plane of 2 collapsing stars (Model III[a] and
[b] of Table \ref{tbl:initial}).  Snapshots are plotted at the parameter 
($t/t_{\rm dyn}$, $\rho^{*}_{\rm max}$) $=$ 
III(a)-i ($3.19$, $2.70 \times 10^{-4}$), 
III(a)-ii ($7.77$, $4.02 \times 10^{-3}$), 
III(b)-i ($3.20$, $3.28 \times 10^{-4}$), 
III(b)-ii ($7.77$, $1.92 \times 10^{-3}$), respectively.
The contour lines denote coordinate densities
$\rho^{*} = \rho^{*}_{\rm max} \times 10^{- 0.220 (16-i)}  (i=1, \cdots, 15)$.
\label{fig:n25con}
}
\end{figure*}

We show the quadrupole diagnostic $Q$ throughout the evolution in Fig. 
\ref{fig:dig}.  We determine that the system is stable to $m=2$ mode when the 
quadrupole diagnostic remains small throughout the evolution.  We also 
determine that the system is unstable when the diagnostic grows exponentially 
at its early stage of evolution.  It is clearly shown in Fig. \ref{fig:dig} 
that the star is more unstable to the bar mode when the structure of the star 
is toroidal at its equilibrium stage than when it is spheroidal.  Even in a 
spheroidal case, the dynamical instability sets in after the core bounce in 
case of model I(b) (see Table \ref{tbl:initial}).  Note also that the 
diagnostic damps out especially for the soft equation of state ($n=2$, 
$2.5$).  The above effect corresponds to the destruction of the toroidal 
structure of the star (see Fig. \ref{fig:density}).

We show the density contour in the intermediate stage and in the final stage 
of the rotational core collapse in Figs. \ref{fig:n15con} -- \ref{fig:n25con}.
In the intermediate stage of the rotational core collapse, the bar 
instability grows due to the nonaxisymmetric instability.  The deformation 
rate of the bar is significantly higher for the case of a toroidal star at 
equilibrium than that of a spheroidal star.  At the termination of integration,
the equatorial shape of the star is almost spherical except for the case of 
model I(a).  The coincidence of the destruction of the toroidal structure 
in Fig. \ref{fig:density} at the termination of integration indicates that 
the structure of torus plays a significant role in enhancing bar instability.

\begin{figure*}
\begin{center}
\includegraphics[width=0.7\textwidth]{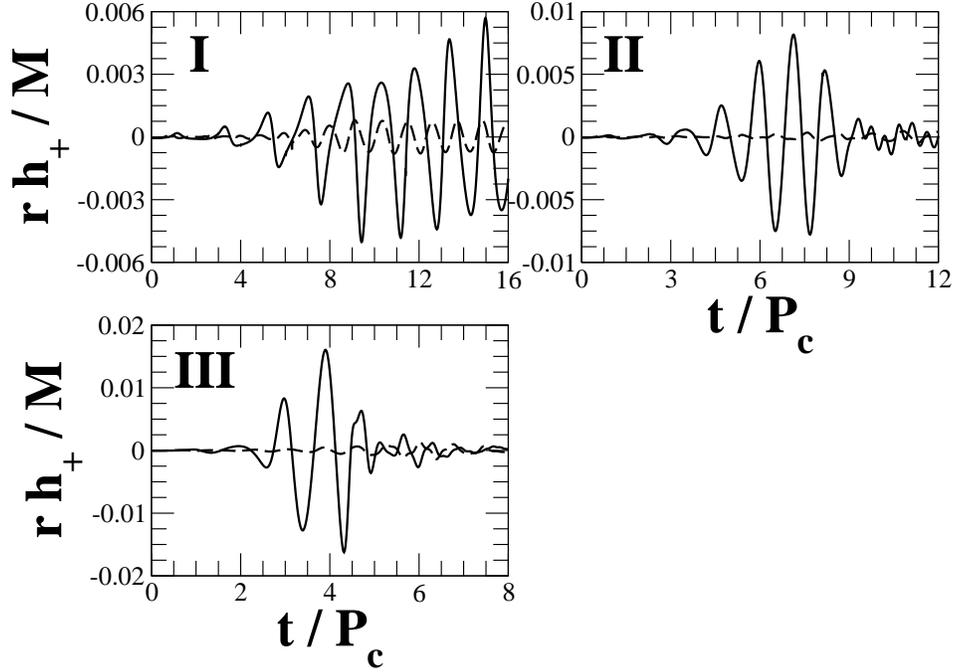}
\end{center}
\caption{
Gravitational waveforms from a distant observer along the rotational axis of 
the equilibrium star.  Roman character in the panel corresponds to the model 
type in Table \ref{tbl:initial}.  Solid lines and dash lines represent 
toroidal and spheroidal stars, which correspond to $a$ and $b$ in Table 
\ref{tbl:initial}, respectively.
\label{fig:gw}
}
\end{figure*}

We show the gravitational waveforms along the observer in the rotational axis
in Fig. \ref{fig:gw}.  The gravitational waves amplify during the bar 
formation.  Since the dynamical bar plays an important role in our case of 
rotational core collapse, our result is quite different from the previous 
picture of the 3D Newtonian results \citep{RMR}.  In fact, the amplitude 
in our calculation observed from the rotational axis grows about 20 times 
larger than that of core bounce which has a peak in the waveform.  Note also 
that the quasiperiodic waves retain several rotation periods for stiff 
equation of state, due to the persistence of the toroidal structure.

\begin{figure*}
\begin{center}
\includegraphics[width=0.7\textwidth]{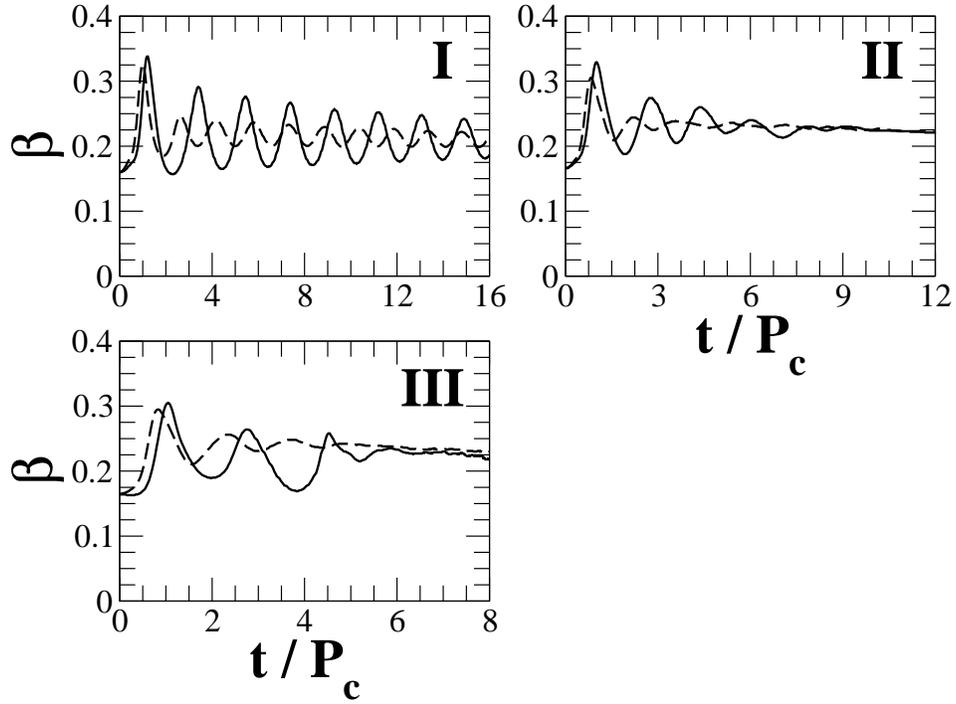}
\end{center}
\caption{
Diagnostics of dynamical bar instability $\beta$ as a function of time.  Roman 
character in the panel corresponds to the model type in 
Table~\ref{tbl:initial}.  Solid lines and dash lines represent toroidal and 
spheroidal star, which corresponds to $a$ and $b$ in Table~\ref{tbl:initial}.
\label{fig:tw}
}
\end{figure*}

\begin{figure*}
\begin{center}
\includegraphics[width=0.7\textwidth]{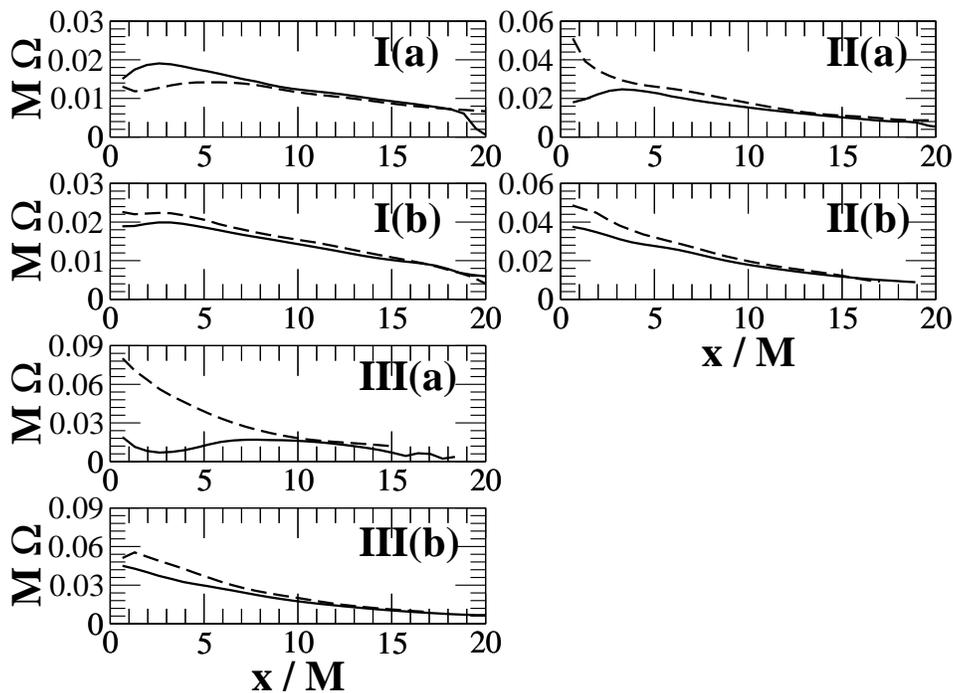}
\end{center}
\caption{
Snapshot of the orbital angular velocity along the $x$-axis.
Solid lines and dash lines represent $t/P_{\rm c}=$ 
  I(a) (9.61, 16.0),   I(b) (9.67, 15.3), 
 II(a) (2.39, 12.0),  II(b) (3.25, 12.0), 
III(a) (3.19, 7.77), III(b) (3.20, 7.77), respectively.
\label{fig:omg}
}
\end{figure*}

We also show $\beta$ in the evolution in Fig. \ref{fig:tw}, which is regarded 
as a diagnostic of the dynamical bar instability in the equilibrium star.  
Although $\beta$ behaves quite similar in the different polytropic index, the 
bar structure persists for at least several rotation periods in case of 
the model I(a), which corresponds to the persistence of the toroidal structure
(see Fig \ref{fig:density}).  We also show the distribution of the angular 
velocity in the intermediate stage and at the termination of the integration 
in Fig. \ref{fig:omg}.  A sharp dip at the center in the angular velocity of 
model I(a) at $t=9.61 P_{\rm c}$ potentially has a redistribution of the 
angular momentum.

\section{Discussion}
\label{sec:Discussion}
We investigate the role of dynamical bar instability in rotational core 
collapse by means of hydrodynamic simulations in conformally flat 
approximation in general relativity.  We specifically focus on the 
structure of the star to see whether it takes a significant role in enhancing  
dynamical bar instability.

We find that the structure of the star takes a significant role in enhancing 
the dynamical bar instability at core bounce.  Since the angular velocity of 
the collapsing star has already reached the maximum inside a certain 
cylindrical radius to produce a toroidal structure, the angular momentum can 
only shift outward at the bounce.  For a spheroidal star, the angular 
momentum can shift both inward and outward at bounce since it does not reach 
the ``Keplarian''.  This means that rotational core collapse for the 
spheroidal case cannot significantly break the central core of the star.  
Consequently, in case of a toroidal star a bar structure is easily 
constructed during the evolution.  Note that for a soft equation of state 
($n=2.0$, $2.5$) the amplitude of gravitational waves decreases, as the torus 
is destroyed in the rotational core collapse.  Therefore the torus is the key 
issue to trigger the bar formation.

Once a bar has formed, the amplitude of gravitational waves has significantly 
increased due to the nonaxisymmetric deformation of the star.  The previous 2D 
calculation shows that a peak amplitude of gravitational waves comes from the 
core bounce of the star, and its behavior coincides with that of 3D 
calculation.  In our results, gravitational radiation is dominantly generated 
in the bar formation process.  Our results of the soft equation of state 
qualitatively agree with the results in full general relativity \citep{SS05}.  
Thus, the characteristic amplitude and frequency of gravitational waves in 
the collapsing star can be written as
\begin{eqnarray}
f_{\rm gw} &\approx& \frac{1}{2 \pi t_{\rm dyn}} =
100~[{\rm Hz}] \left( \frac{M_{\odot}}{M} \right)
\left( \frac{40M}{R} \right)^{3/2}
,\\
h_{\rm gw} &\approx& 4.8 \times 10^{-23}
\left( \frac{M}{M_{\odot}} \right)
\left( \frac{10 {\rm M pc}}{d} \right)
\left( \frac{rh/M}{0.01} \right).
\nonumber
\\
\end{eqnarray}
Therefore, gravitational waves from a dynamical bar in a collapsing neutron 
star will be detected in advanced LIGO.

\begin{table}
\caption
{Magnitude of higher order correction for the central 
lapse at bounce}
\begin{center}
\begin{tabular}{c c c c c}
\hline
\hline
Model &
${t_{\rm bce}/P_{\rm c}}$\footnotemark[10] &
$\alpha_{\rm c}$\footnotemark[11] &
$M_{\rm src}/R_{\rm src}$\footnotemark[12] &
$(M_{\rm src}/R_{\rm src})^3 [\%]$
\\
\hline
I-a & $1.22$ & $0.872$ & $0.137$ & $0.257$
\\
I-b & $1.00$ & $0.848$ & $0.166$ & $0.457$
\\
\hline
II-a & $1.05$ & $0.843$ & $0.172$ & $0.509$
\\
II-b & $0.88$ & $0.806$ & $0.218$ & $1.04$
\\
\hline
III-a & $1.08$ & $0.835$ & $0.181$ & $0.593$
\\
III-b & $0.90$ & $0.803$ & $0.222$ & $1.09$
\\
\hline
\end{tabular}
\label{tbl:lapse}
\footnotetext[10]{$t_{\rm bce}$: Bounce time}
\footnotetext[11]{Central lapse}
\footnotetext[12]{$M_{\rm src}$: Characteristic mass of the source; 
$R_{\rm src}$: Characteristic radius of the source}
\end{center}
\end{table}

Finally we mention the validity of the conformally flat approximation 
in our 6 different collapsing stars.  The approximation has two issues: 
that it contains the first post-Newtonian order of general relativity, 
and that it neglects gravitational radiation.  First, we investigate the 
central lapse $\alpha_{\rm c}$ at core bounce to check whether the first
post-Newtonian order of general relativity is reasonable approximation in 
our model.  The central lapse can be expanded by the compaction 
(characteristic mass $M_{\rm src}$ and radius $R_{\rm src}$) of the source 
[Eq. (19.13) of Ref. \citep{MTW}] as
\begin{equation}
\alpha_{\rm c} = 
1 - \frac{M_{\rm src}}{R_{\rm src}} 
+ \frac{1}{2} \left( \frac{M_{\rm src}}{R_{\rm src}} \right)^{2}
+ o \left( \left( \frac{M_{\rm src}}{R_{\rm src}} \right)^{3} \right).
\label{eqn:lapsec}
\end{equation}
Note that the shift appears in the $g_{tt}$ component of the 4-metric at the 
second post-Newtonian order of general relativity.  All models of our 6 
collapsing stars bounce at $\alpha_{\rm c} \gtrsim 0.8$, and therefore the 
value of central lapse corresponds to the compaction of the source 
$M_{\rm src} / R_{\rm src} \lesssim 0.22$ from Eq. (\ref{eqn:lapsec}).  
For each model, we summarize the magnitude of higher order correction 
[$(R_{\rm src}/M_{\rm src})^{3}$ term] from the first post-Newtonian order 
in Table \ref{tbl:lapse}.  Note that the central lapse at bounce is the 
minimal one throughout the evolution.  Since the magnitude of the higher 
order correction is $(M_{\rm src} / R_{\rm src})^{3} \lesssim 0.011$, we can 
roughly claim that the first post-Newtonian gravity is a satisfactory 
approximation to describe the full general relativity in the relative error 
rate of few percent in our calculation, depending on the coefficient of the 
higher order term.  Second, it is also a quite satisfactory approximation 
to neglect gravitational radiation in our dynamics, since our main target in 
this paper is to focus on the enhancement of the dynamical bar instabilities 
which occur in dynamical time scale.  Gravitational waves affect the whole 
dynamics in secular time scale, which is much longer than the dynamical 
time scale, and hence we can safely neglect such effect in this paper.  
Therefore, conformally flat approximation is a satisfactory one in our 
computation.

\acknowledgments
We would like to thank Thomas Janka and Ewald M\"uller for their kind 
hospitality at Max-Plank-Institut f\"ur Astrophysik, where part of this 
work was done.  We thank Takashi Nakamura for his continuous encouragement 
and valuable advice.  We also thank Yukiya Aoyama and Jun Nakano for their 
constructive comments on parallelizing my code.  This work was supported in 
part by MEXT Grant-in-Aid for young scientists (No. 200200927).  Numerical 
computations were performed on the VPP-5000 machine in the Astronomical Data 
Analysis Center, National Astronomical Observatory of Japan, on the SGI Origin 
3000 machine in the Yukawa Institute for Theoretical Physics, Kyoto 
University, on the FUJITSU HPC2500 machine in the Academic Center for 
Computing and Media Studies, Kyoto University, and on the Pentium-4 cluster 
machine in the Theoretical Astrophysics Group at Kyoto University.



\begin{thebibliography}{99}
\bibitem[Cutler and Thorne(2002)]{CT02}
C.~Cutler and K.~S.~Thorne,
in {\it Proceedings of the 16th International Conference of General 
Relativity and Gravitation}
edited by N.~T.~Bishop and S.~D.~Maharaj 
(World Scientific, New Jersey, 2002), p. 72.
%
\bibitem[Fryer and Heger(2000)]{FH00}
C.~L.~Fryer, A.~Heger,
\apj {\bf 541}, 1033 (2000).
%
\bibitem[Kotake et al.(2003)]{KYS03}
K.~Kotake, S.~Yamada, and K.~Sato,
\prd {\bf 68}, 044023 (2003).
%
\bibitem[M\"{u}ller, Rampp, Buras, Janka, and Shoemaker(2004)]{MRBJS}
E.~M\"{u}ller, M.~Rampp, R.~Buras, H.-T.~Janka, D.~H.~Shoemaker,
\apj {\bf 603}, 221 (2004).
%
\bibitem[Dimmelmeier, Font, and M\"{u}ller(2001)]{DFM01}
H.~Dimmelmeier, J.~A.~Font, and E.~M\"{u}ller,
\apj  {\bf 560}, L163 (2001).
%
\bibitem[Dimmelmeier, Font, and M\"{u}ller(2002)]{DFM02}
H.~Dimmelmeier, J.~A.~Font, and E.~M\"{u}ller,
Astron.\ Astrophys.\  {\bf 388}, 917 (2002).
%
\bibitem[Cerda-Duran et al.(2004)]{DFDFIMS}
P.~Cerda-Duran, G.~Faye, H.~Dimmelmeier, J.~A.~Font, J.~M.~Ibanez, 
E.~M\"{u}ller, G.~Sch\"{a}fer,
astro-ph/0412611 [Astron. Astrophys. (to be published)].
%
\bibitem[Tohline, Durisen, and McCollough(1985)]{TDM}
J.~E.~Tohline, R.~H.~Durisen and M.~McCollough, 
\apj {\bf 298}, 220 (1985).
%
\bibitem[Durisen et al.(1986)]{DGTB}
R.~H.~Durisen, R.~A.~Gingold, J.~E.~Tohline, and A.~P.~Boss, 
\apj {\bf 305}, 281 (1986).
%
\bibitem[Williams and Tohline(1988)]{WT}
H.~A.~Williams and J.~E.~Tohline,
\apj {\bf 334}, 449 (1988).
%
\bibitem[Houser, Centrella, and Smith(1994)]{HCS}
J.~L.~Houser, J.~M.~Centrella and S.~C.~Smith, 
\prl {\bf 72}, 1314 (1994).
%
\bibitem[Smith, Houser, and Centrella(1996)]{SHC}
S.~C.~Smith, J.~L.~Houser, and J.~M.~Centrella,
\apj {\bf 458}, 236 (1996).
%
\bibitem[Houser and Centrella(1996)]{HC}
J.~L.~Houser and J.~M.~Centrella, 
\prd {\bf 54}, 7278 (1996).
%
\bibitem[Toman et al.(1998)]{TIPD}
J.~Toman, J.~N.~Imamura, B.~J.~Pickett, and R.~H.~Durisen, 
\apj {\bf 497}, 370 (1998).
%
\bibitem[New, Centrella, and Tohline(2000)]{NCT}
K.~C.~B. New, J.~M.~Centrella, and J.~E.~Tohline, 
\prd {\bf 62}, 064019 (2000).
%
\bibitem[Liu and Lindblom(2001)]{LL}
Y.-T.~Liu and L.~Lindblom,
Mon. Not. R. Astron. Soc. {\bf 324}, 1063 (2001).
%
\bibitem[Liu(2002)]{Liu02}
Y.-T.~Liu, 
\prd {\bf 65}, 124003 (2002).
%
\bibitem[Tohline and Hachisu(1990)]{TH90}
J.~E.~Tohline, and I.~Hachisu, 
\apj {\bf 361}, 394 (1990). 
%
\bibitem[Pickett, Durisen, and Davis(1996)]{PDD}
B.~K.~Pickett, R.~H.~Durisen, and G.~A.~Davis, 
\apj {\bf 458}, 714 (1996).
%
\bibitem[Shibata, Karino, and Eriguchi(2002)]{SKE}
M.~Shibata, S.~Karino, and Y.~Eriguchi, 
Mon. Not. R. Astron. Soc. {\bf 334}, L27 (2002);
{\bf 343}, 619 (2003).
%
\bibitem[Shibata, Baumgarte, and Shapiro(2000)]{SBS00}
M.~Shibata, T.~W.~Baumgarte, and S.~L.~Shapiro, 
\apj {\bf 542}, 453 (2000).
%
\bibitem[Saijo et al.(2001)]{SSBS01}
M.~Saijo, M.~Shibata, T.~W.~Baumgarte, and S.~L.~Shapiro, 
\apj {\bf 548}, 919 (2001).
%
\bibitem[Zwerger and M\"{u}ller(1997)]{ZM}
T.~Zwerger and E.~M\"{u}ller,
Astron. Astrophys. {\bf 320}, 209 (1997).
%
\bibitem[Siebel et al.(2003)]{Siebel03}
F.~Siebel, J.~A.~Font, E.~M\"{u}ller, and P.~Papadopoulos,
\prd {\bf 67}, 124018 (2003).
%
\bibitem{Shibata04}
M.~Shibata,
\prd {\bf 67}, 024033 (2003).
%
\bibitem[Shibata and Sekiguchi(2004)]{SS04}
M.~Shibata and Y.-I.~Sekiguchi,
\prd {\bf 69}, 084024 (2004).
%
\bibitem[Stergioulas, Apostolatos, and Font(2004)]{SAF}
N.~Stergioulas, T.~A.~Apostolatos, J.~A.~Font,
Mon. Not. R. Astron. Soc. {\bf 352}, 1089 (2004).
%
\bibitem[Fryer, Holz and Hughes(2002)]{FHH02}
C.~L.~Fryer, D.~E.~Holz, and S.~A.~Hughes,
\apj {\bf 565}, 430 (2002).
%
\bibitem[Duez, Shapiro, and Yo(2004)]{DSY}
M.~D.~Duez, S.~L.~Shapiro, and H.-J.~Yo, 
\prd {\bf 69}, 104016 (2004).
%
\bibitem[Shibata and Sekiguchi(2005)]{SS05}
M.~Shibata and Y.-I.~Sekiguchi,
\prd {\bf 71}, 024014 (2005).
%
\bibitem[Zink et al.(2005)]{Zink05}
B.~Zink, N.~Stergioulas, I.~Hawke, C.~D.~Ott, 
E.~Schnetter, and E.~M\"{u}ller,
gr-qc/0501080.
%
\bibitem[Brown(2001)]{Brown}
J.~D.~Brown, 
in {\it Astrophysical Sources for Ground-based Gravitational Wave
Detectors}, edited by J.~M.~Centrella (American Institute of
Physics, New York, 2001), p. 234.
%
\bibitem[Rampp, M\"{u}ller, and Ruffert(1998)]{RMR}
M.~Rampp, E.~M\"{u}ller, and M.~Ruffert, 
Astron. Astrophys. {\bf 332}, 969 (1998).
%
\bibitem[Ott et al.(2004)]{Ott04}
C.~D.~Ott, A.~Burrows, E.~Livne, and R.~Walder,
\apj {\bf 600}, 834 (2004).
%
\bibitem[Shibata and Uryu(2000)]{SU00}
M.~Shibata, and K.~Ury\={u},
\prd {\bf 61}, 064001 (2000).
%
\bibitem[Shibata and Uryu(2002)]{SU02}
M.~Shibata, and K.~Ury\={u},
Prog. Theor. Phys. {\bf 107}, 265 (2002).
%
\bibitem[Shibata, Taniguchi, and Ury\={u}(2003)]{STU03}
M.~Shibata, K.~Taniguchi, K.~Ury\={u},
\prd {\bf 68}, 084020 (2003).
%
\bibitem[Isenberg and Nester(1980)]{IN}
J.~Isenberg and J.~Nester, 
in {\it General Relativity and Gravitation Vol. 1: one hundred years
after the birth of Albert Einstein},
edited A.~Held (Plenum Press, New York, 1980), p. 23.
%
\bibitem[Wilson and Mathews(1989)]{WM89}
J.~R.~Wilson and G.~J.~Mathews, 
in {\it Frontiers in numerical relativity},
edited by C.~R.~Evans, L.~S.~Finn, D.~W.~Hobill
(Cambridge Univ. Press, Cambridge, 1989), p. 306.
%
\bibitem[Saijo(2004)]{Saijo04}
M.~Saijo, 
\apj {\bf 615}, 866 (2004).
%
\bibitem[Chandrasekhar(1965)]{Chandra65}
S.~Chandrasekhar, 
\apj {\bf 142}, 1488 (1965).
%
\bibitem[Blanchet, Damour, and Sch\"{a}fer(1990)]{BDS}
L.~Blanchet, T.~Damour, and G.~Sch\"{a}fer, 
Mon. Not. R. Astron. Soc. {\bf 242}, 289 (1990).
%
\bibitem[Murata, Natori, and Karaki(1990)]{PCG}
K.~Murata, R.~Natori,  and Y.~Karaki,
Large-scale numerical simulation (Iwanami, Tokyo, 1990), 
Chap. 5.2 (in Japanese).
%
\bibitem[Cook, Shapiro, and Teukolsky(1992)]{CST92}
G.~B.~Cook, S.~L.~Shapiro, and S.~A.~Teukolsky, 
\apj {\bf 398}, 203 (1992).
%
\bibitem[Komatsu, Eriguchi, and Hachisu(1989)]{KEH89}
H.~Komatsu, Y.~Eriguchi, and I.~Hachisu, 
Mon. Not. R. Astron. Soc. {\bf 237}, 355 (1989).
%
\bibitem[Saijo, Baumgarte, and Shapiro(2003)]{SBS03}
M.~Saijo, T.~W.~Baumgarte, and S.~L.~Shapiro, 
\apj {\bf 595}, 352 (2003).
%
\bibitem[Misner, Thorne, and Wheeler(1973)]{MTW}
C.~W.~Misner, K.~S.~Thorne, and J.~A.~Wheeler,
Gravitation (W. H. Freeman and Company, New York, 1973), 
Sec. 19.3.
%
\end{thebibliography}
\end{document}